\documentclass[twocolumn]{aastex63}

\accepted{}
\submitjournal{AJ}

\shorttitle{The Nature of 2M1155$-$79B}
\shortauthors{Dickson-Vandervelde et al.}
\graphicspath{{./}{figures/}}

\begin{document}

\title{Investigating 2M1155-7919B: a Nearby, Young, Low-Mass Star Actively Accreting from a Nearly Edge-on, Dusty Disk}

\correspondingauthor{D. Annie Dickson-Vandervelde}
\email{dad1197@rit.edu}

\author[0000-0002-4555-5144]{D. Annie Dickson-Vandervelde}
\affiliation{School of Physics and Astronomy, 
  Rochester Institute of Technology, Rochester NY 14623, USA}
  \affiliation{Laboratory for Multiwavelength Astrophysics, Rochester Institute of Technology}

\author[0000-0002-3138-8250]{Joel H. Kastner}
\affiliation{School of Physics and Astronomy, 
  Rochester Institute of Technology, Rochester NY 14623, USA}
 \affiliation{Laboratory for Multiwavelength Astrophysics, Rochester Institute of Technology}
\affiliation{Center for Imaging Science, 
  Rochester Institute of Technology; jhk@cis.rit.edu}

\author[0000-0002-2592-9612]{Jonathan Gagn\'e}
\affiliation{Plan\'etarium Rio Tinto Alcan, Espace pour la Vie, 4801 av. Pierre-de Coubertin, Montr\'eal, Qu\'ebec, Canada}
\affiliation{Institute for Research on Exoplanets, Universit\'e de Montr\'eal, D\'epartement de Physique, C.P.~6128 Succ. Centre-ville, Montr\'eal, QC H3C~3J7, Canada}

\author[0000-0002-6294-5937]{Adam C. Schneider}
\affil{United States Naval Observatory, Flagstaff Station, 10391 West Naval Observatory Rd., Flagstaff, AZ 86005, USA}
\affil{Department of Physics and Astronomy, George Mason University, MS3F3, 4400 University Drive, Fairfax, VA 22030, USA}

\author[0000-0001-6251-0573]{Jacqueline Faherty}
\affiliation{Department of Astrophysics, American Museum of Natural History, Central Park West at 79th St., New York, NY 10024, USA}

\author[0000-0002-3947-5586]{Emily C. Wilson}
\affiliation{School of Physics and Astronomy, 
  Rochester Institute of Technology, Rochester NY 14623, USA}
\affiliation{Center for Computational Relativity and Gravitation, Rochester Institute of Technology}

\author[0000-0001-5907-5179]{Christophe Pinte}
\affiliation{Monash Centre for Astrophysics (MoCA) and School of Physics and Astronomy,
Monash University, Clayton, Vic 3800, Austrailia}

\author[0000-0002-1637-7393]{Francois M\'enard}
\affiliation{Univ. Grenoble Alpes, CNRS, IPAG, F-38000 Grenoble, France}

\begin{abstract}
We investigate the nature of an unusually faint member of the $\epsilon$ Cha Association ($D\sim100$ pc, age $\sim5$ Myr), the nearest region of star formation of age $<$8 Myr. This object, 2MASS J11550336-7919147 (2M1155$-$79B), is a wide ($\sim$580 AU) separation, comoving companion to low-mass (M3) $\epsilon$ Cha Association member 2MASS J11550485-7919108 (2M1155$-$79A). We present near-infrared spectra of both components, along with analysis of photometry from Gaia EDR3, 2MASS, VHS, and WISE. The near-IR spectrum of 2M1155$-$79B displays strong He {\sc i} 1.083 emission, a sign of active accretion and/or accretion-driven winds from a circumstellar disk. Analysis of WISE archival data reveals that the mid-infrared excess previously associated with 2M1155$-$79A instead originates from the disk surrounding 2M1155$-$79B. Based on these results, as well as radiative transfer modeling of its optical/IR spectral energy distribution, we conclude that 2M1155$-$79B is most likely a young, late-M, star that is partially obscured by, and actively accreting from, a nearly edge-on circumstellar disk. This would place 2M1155$-$79B among the rare group of nearby ($D\lesssim100$ pc), young (age $<$10 Myr) mid-M stars that are orbited by and accreting from highly inclined protoplanetary disks. Like these systems, the 2M1155$-$79B system is a particularly promising subject for studies of star and planet formation around low-mass stars.
\end{abstract}


\section{Introduction} \label{sec:intro}

\begin{deluxetable*}{lrrrrrrrrrrr}
\tablecaption{\sc 2M1155$-$79AB: Gaia EDR3 Astronomy and Photometry}
\tabletypesize{\scriptsize}
\label{table:data}
\tablehead{\colhead{Name}  & \colhead{RA} & \colhead{Dec} & \colhead{$\pi$} & \colhead{PMRA} & \colhead{PMDec} & \colhead{G} &  \colhead{G$-$G$_{RP}$} & \colhead{RUWE}\\ \colhead{}& \colhead{(deg)} & \colhead{(deg)} & \colhead{(mas)} & \colhead{(mas/yr)} & \colhead{(mas/yr)} & \colhead{(mag)} & \colhead{(mag)}}
\startdata
2M1155$-$7919A & 178.769132 & -79.319756 & 9.81$\pm$0.03 & -41.35$\pm$0.03 & -4.56$\pm$0.031 & 14.803 & 1.333 & 1.196\\
2M1155$-$7919B & 178.762736 & -79.320829 & 9.49$\pm$0.43 & -41.59$\pm$0.55 & -4.63$\pm$0.51 & 19.954 & 1.396 & 1.104\\
\enddata
\end{deluxetable*}

Nearby associations of young, comoving stars (generally referred to as Nearby Young Moving Groups; hereafter NYMGs) are prime candidates for studies of the early evolution of low-mass pre-main sequence (pre-MS) stars, juvenile brown dwarfs, and newly formed planets \citep{IAUProceedings2016,GagneFaherty2018}. In recent years there have been a handful of identifications of very wide ($\sim$100--1000 AU projected separation) binaries consisting of young stars and substellar objects in these NYMGs, as well as in nearby star formation regions \citep[HD 106906 b, 1RXS 160929.1-210524 B, CT Cha B, and DENIS-P J1538317-103850;][]{Wu2015a,Wu2015b,Wu2016,N-T2020}.  

The faint 2MASS source J11550336-79191147 (henceforth 2M1155$-$79B) is a curious young, low-mass object that was discovered via a Gaia Data Release 2 \citep[DR2,][]{Gaia2016} search for wide, comoving companions to known members of the $\epsilon$ Cha Association ($\epsilon$CA; $D\sim100$ pc, age $\sim$5 Myr), which represents the nearest region of star formation of age $<$8 Myr \citep{DickVand2020,DickVand2021}. 2M1155$-$79B is the companion to 2MASS J11550486-7919108 (henceforth 2M1155$-$79A) with a projected separation of 5$"$.75, equivalent to 582 AU at PA 227.9$^\circ$ \citep{Murphy2013,DickVand2020}. Astrometry from Gaia Early Data Release 3 (EDR3) confirms that 2M1155$-$79A and 2M1155$-$79B are equidistant \citep[$D = $101.4$\pm$0.3 pc,][]{BailerJones2021} and comoving, to within the uncertainties (Table 1). The M3 type star 2M1155$-$79A was previously known as T Cha~B, following its identification as an apparent very wide-separation comoving companion to T Cha, host to one of the nearest known examples of a highly inclined protoplanetary disk \citep{Kastner2012}.  However, Gaia DR2 subsequently revealed small but statistically significant differences between the parallaxes and proper motions of T Cha and 2M1155$-$79A \citep[][]{Kastner2018}\footnote{Gaia EDR3 data appear to confirm that T Cha and 2M1155$-$79A are neither equidistant nor comoving (within the uncertainties).}.

Using substellar object population properties, \citet{DickVand2020} found that the J magnitude and J-H color of 2M1155$-$79B corresponded to a spectral type of M9/L0, which would place it just below the boundary between brown dwarfs and massive planets assuming its age is 5 Myr. This mass estimate would place 2M1155$-$79B in the interesting position of a planet orbiting its host (2M1155$-$79A) at a projected semimajor axis of $\sim$600 AU --- similar to the projected separations of the aforementioned substellar-object-hosting systems HD 106906, 1RXS 160929.1-210524, CT Cha, and DENIS-P J1538317-103850. The formation mechanisms of such wide-orbit massive planets remain subject to debate; proposed methods of formation range from dynamical interactions to standard star formation mechanisms \citep[e.g.,][]{Lagrange2016,Rodet2019,Lodieu2021}

In this paper, we present near-infrared (NIR) spectroscopy of the 2M1155$-$79AB system along with analysis of available Gaia EDR3, 2MASS, and WISE photometry. The results suggest a different nature for 2M1155$-$79B: namely, that it is a low-mass pre-MS star partially occulted by, and actively accreting from, a highly inclined, dusty disk. In Section~\ref{sec:obs}, we describe the NIR spectroscopic observations. In Section~\ref{sec:results}, we present the results of these observations along with analysis of archival photometry of the 2M1155$-$79AB system. In Section~\ref{sec:discussion}, we discuss the evidence for a disk around 2M1155$-$79B and compare the object to 2M1155$-$79A. We also compare the 2M1155$-$79B star-disk system to a set of potentially analogous very low-mass star-disk systems:  fellow $\epsilon$CA member 2MASS J12014343-7835472, member of the 8 Myr TW Hys Association TWA 30B, and the 1-5 Myr-old DENIS-P J1538317-103850. 

\begin{figure*}
    \centering
    \includegraphics[width=0.95\linewidth]{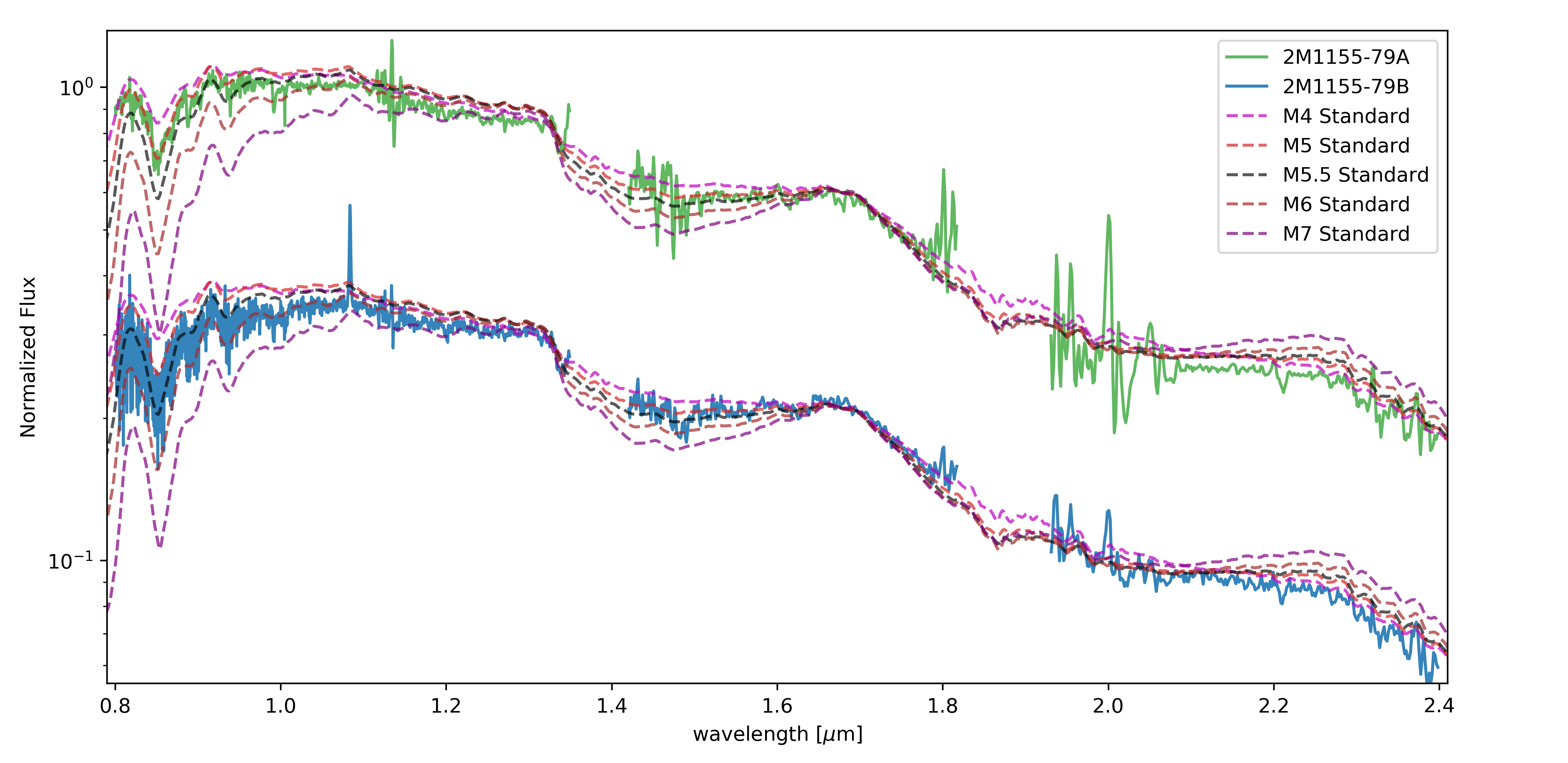}
    \caption{Near-IR spectra obtained with the FIRE instrument on Magellan for 2M1155$-$79A (green) and 2M1155$-$79B (blue). The fluxes of both FIRE spectra have been normalized and a constant offset applied to 2M1155$-$79B, for purposes of visualization. Five spectra from the \citet{Luhman2017} Young Star Spectral Library, M4 (magenta) M5 (red), M5.5 (black), M6 (brown), and M7 (purple), are also displayed with intensities adjusted to match the FIRE spectra at 1.7 $\mu$m.}
    \label{fig:FIREspec}
\end{figure*}

\section{Observations}\label{sec:obs}

Near-IR spectra were obtained on February 12 and 13, 2020 (hereafter Night 1 and Night 2, respectively) using the FIRE instrument on the 6.5 m Magellan telescope. FIRE spectra were obtained in prism mode, which has a resolving power of $\sim$450 across the 0.8--2.5 $\mu$m spectral range. Observations of 2M1155$-$79B were obtained both nights and observations of 2M1155$-$79A were obtained on Night 2 (Fig.~\ref{fig:FIREspec}). 

The FIRE data were reduced using a custom version of the FIREHOSE pipeline based on the MASE pipeline \citep{Bochanski2009} written in the Interactive Data Language (IDL)\footnote{Available at \url{https://github.com/jgagneastro/FireHose_v2}}. This modified version includes subroutines of the SpexTool package \citep{Vacca2003,Cushing2004} to facilitate the rejection of bad pixels and detector hot spots when the 1D extracted spectra of individual exposures. All raw exposures were flat-fielded, and then extracted using the optimal extraction algorithm included in FIREHOSE, and finally combined using the default Robust Weighted Mean option (Robust threshold = 8.0). We used Neon-Argon lamp exposures to determine the wavelength calibration. The signal-to-noise per resolution element is approximately 180 (70 per pixel) for 2M1155$-$79B and about 260 (100 per pixel) for 2M1155$-$79A.

While the overall shape and features of the spectra for 2M1155$-$79B did not change between Night 1 and Night 2, the spectrum obtained on Night 1 displays lower signal-to-noise ratio (SNR), especially in the $\sim$2 $\mu$m region. Hence, the analysis and discussion in this paper refers to the Night 2 spectrum, unless otherwise specified.

The majority of $\epsilon$CA members show no significant intervening reddening \citep{Murphy2013}, consistent with the general lack of extinction in the direction of the group ($E(B-V) < 0.03$) out to distances of $\sim$160 pc \citep{Lallement2019}. However, the 2M1155$-$79AB system (like neighboring T~Cha) is seen projected toward the middle of a small dust cloud \citep{Murphy2013,Sacco2014}. Hence, the spectra of both components were dereddened via the python package \texttt{dust\_extinction} using the \citet{Fitzpatrick2019} reddening model. We adopted $E(B-V)= 0.5$ for 2M1155$-$79A \citep{Murphy2013}, which equates to $A_J \approx 0.44$, 
assuming the standard value of $R_V= 3.1$ \citep[e.g.,][]{Cardelli1989}. Note that this dereddening process does not account for other sources of spectral distortion, such as obscuration or scattering by circumstellar disk material.

\section{Results}\label{sec:results}

\subsection{NIR Spectra}\label{subsec:spec}

\begin{figure}[t]
    \centering
    \includegraphics[width=0.98\linewidth]{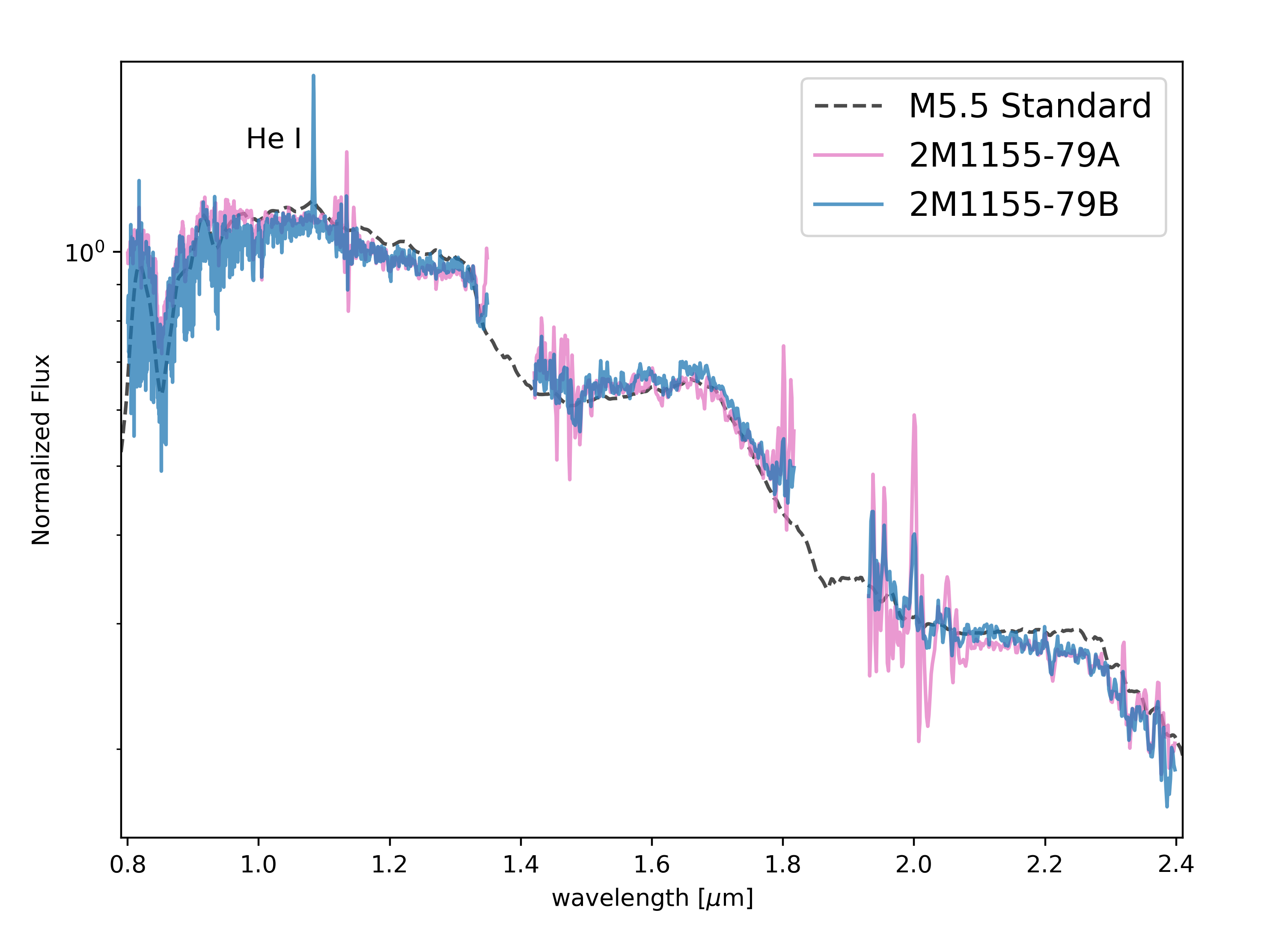}
    \caption{Near-IR spectra of 2M1155$-$79A (green), 2M1155$-$79B (blue), and a M5.5 standard (dashed black), displayed with fluxes normalized to the value at 0.85 $\mu$m. Aside from the 1.083 $\mu$m He {\sc i} emission line seen in 2M1155$-$79B, no other emission lines are detected in either star. Notably, the two companions have very similar morphologies with the exception of the He {\sc i} emission line. }
    \label{fig:ABSpectralComp}
\end{figure}

In Figure~\ref{fig:FIREspec}, we present the dereddened Magellan/FIRE spectra for both 2M1155$-$79A and 2M1155$-$79B overlaid with a range of near-IR spectra of late-M young star spectral standards from \citet{Luhman2017} (henceforth referenced as L17). Before normalizing the two spectra, the 2M1155$-$79A spectrum is on average 10.2$\times$ brighter than its companion. After accounting for this factor of 10 difference in flux, the spectral shapes of 2M1155$-79$A and 2M1155$-$79B are very similar, with some small differences in the depth of the 0.8 $\mu$m absorption feature that could be due to noise (see the two spectra overlaid in Fig.~\ref{fig:ABSpectralComp}). The 1.4$-$1.8 $\mu$m region of the spectrum of 2M1155$-$79B also potentially shows a larger bump than that of 2M1155$-$79A, and 2M1155$-$79B appears slightly redder than 2M1155$-$79B in the 0.8--1.1 $\mu$m region, indicative of a slightly later spectral type. Notably, the spectrum of 2M1155$-$79B displays a strong 1.083 $\mu$m He~{\sc i} emission feature that is absent from the spectrum of 2M1155$-$79A. 

The overall strong similarity of the Magellan/FIRE spectrum of 2M1155$-$79B to that of 2M1155$-$79A indicates that the two components are very similar in spectral type, notwithstanding the factor $\sim$10 smaller spectral flux from 2M1155$-$79B. As is discussed in detail below, this is a surprising result, given that (as noted) 2M1155$-$79A has been classified as M3 \citep{Kastner2012,Murphy2013} whereas 2M1155$-$79B was initially considered a candidate substellar object \citep{DickVand2020}. 
Additionally, while the comparison of the near-IR spectra of the 2M1155$-$79AB paired with those of the L17 standard spectra in Figure~\ref{fig:FIREspec} does leave some ambiguity as to the M spectral subtypes of 2M1155$-$79A and 2M1155$-$79B, it is readily apparent that the near-infrared spectral type of the former component is later than M3. 

The comparison of the overall shapes of the FIRE spectra with those of the L17 young star standard spectra demonstrates that a spectral type of M7 or later can be ruled out; note in particular the flatter slope of both 2M1155$-$79A and B in the 1.4--1.8 and 1.9--2.4 $\mu$m regions relative to the M7 standard. The 2.2 $\mu$m Na {\sc i} absorption line is present in both FIRE spectra, although it is not present in the standards, and the line depth is in agreement with a mid- to late-M spectral type. The 0.8--0.9 $\mu$m range of 2M1155$-$79B has a lower SNR than other regions of the spectrum and so the TiO feature matches with a larger range of spectral types; 2M1155$-$79A is less noisy in that range and more precisely matches to the M5--M6 standard. Again, the slope of the M7 standard is more pronounced than either 2M1155$-$79A or B. Hence, a spectral type in the range of M5--M6 is the best match for both 2M1155$-$79A and B. 

\subsubsection{The He {\sc i} Feature}
Both (Night 1 and 2) spectra of 2M1155$-$79B data show the presence of a strong He {\sc i} 1.083 $\mu$m emission line 
(Figure~\ref{fig:ABSpectralComp}). The emission line is unresolved, placing an upper limit of $\sim$670 km s$^{-1}$ on its velocity width. The equivalent widths (EWs) of the 1.083 $\mu$m emission line, as measured via Gaussian fitting, were $-$11.9 $\pm$ 0.6 \r{A} and $-$12.6 $\pm$ 0.4 \r{A} for the Night 1 and 2 spectra, respectively.

\subsubsection{Low Gravity Features}

We analyzed the 2M1155$-$79B spectrum for potential spectral features that are sensitive to stellar surface gravity and useful for gravity indexing spectral types of M6 and later \citep[][and references therein]{AllersLiu2013}. Both the 1.14 and 2.21 $\mu$m Na {\sc i} absorption lines are visible, consistent with a mid/late-M classification (see above); the Na {\sc i} feature is not present in lower-mass objects. The three FeH features (0.99, 1.20, and 1.55 $\mu$m) are not detected at the current signal-to-noise ratio; this is in agreement with with the morphological spectral type of M5--M6 for 2M1155$-$79B indicated by Figure~\ref{fig:FIREspec} as deeper FeH spectral lines are indicative of later (late-M to early-L) spectral types. 

\subsection{WISE Image Centroids}\label{subsubsec:WISE}

\begin{figure}
    \centering
    \includegraphics[width=0.95\linewidth]{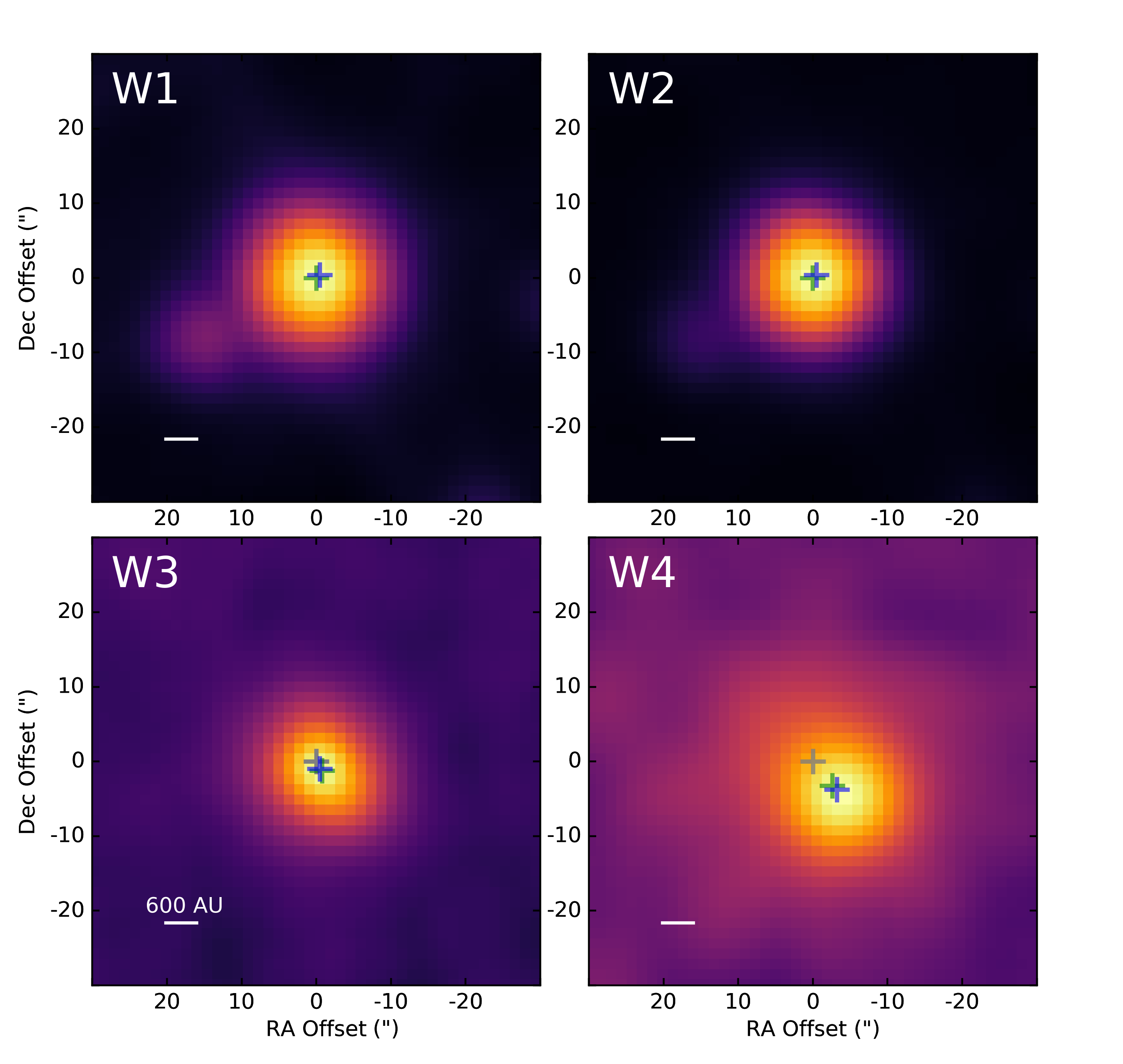}
    \caption{Four WISE archival log-scaled images of the 2M1155$-$79AB system, W1, W2, W3, and W4 (3.6, 4.5, 12, and  22 microns) as labeled. Images are aligned in RA/Dec coordinates, (0,0) being the position of the W1 centroid and are 60$''$x60$''$. The white bar on each image represents the equivalent distance of $\sim$600 AU, the green plus sign marks the center of the PSF as found via a Gaussian, the blue plus sign marks the center of the PSF as found via the peak pixel value, and the grey plus sign in the W3 and W4 images marks the W1 (0,0) position. }
    \label{fig:WISEcent}
\end{figure}

Mid-infrared 3.6, 4.5, 12, and 22 $\mu$m (W1, W2, W3, W4 band) images of the 2M1155$-$79AB binary system obtained by Wide-field Infrared Survey Explorer (WISE) are presented in Fig.~\ref{fig:WISEcent}. The angular resolution of WISE is comparable to the 2M1155$-$79AB system separation ($\sim$6$''$); however, it is apparent that the source centroid shifts from W1 (3.6 $\mu$m) to W4 (22 $\mu$m) by a displacement similar to this angular offset. In order to determine which component dominates each WISE band, we performed a centroid analysis on all four WISE images and calculated the shift of center of the point spread function (PSF). We used two methods to determine the center of the WISE emission: Gaussian PSF centroid fitting and peak-pixel position determination. The two methods yield the same results to within the uncertainties (Fig.~\ref{fig:WISEcent}). Analysis of unWISE data \citep{Schlafly2019} yields similar results. Considering the SED morphologies of 2M1155$-$79A and 2M1155$-$79B (see Sec~\ref{subsubsec:SED}, Fig.~\ref{fig:SED}), the majority of the flux in W1 is presumed to be from 2M1155$-$79A and, ergo, we consider the photocenter found via our analysis of the W1 image to be the position of 2M1155$-$79A. 

The measured offsets with respect to W1 are 0.03" (PA = 270$^\circ$), 1.44" (PA = 242$^\circ$), and 4.14" (PA = 229$^\circ$) in the W2, W3, and W4 band images, respectively. The W4 angular offset and PA, relative to W1, are similar to the projected separation (5.75") and PA (227$^\circ$) of 2M1155$-$79B, relative to 2M1155$-$79A \citep{DickVand2020}. We conclude that 2M1155$-$79A dominates the flux in the shorter WISE bands (W1 and W2), while the contribution from 2M1155$-$79B becomes significant at 12 $\mu$m (W3) and dominates the flux at 22 $\mu$m (W4). Thus, the mid-IR excess due to thermal emission from circumstellar dust --- originally attributed to 2M1155$-$79A \citep{Kastner2012,Murphy2013} --- is in fact associated with 2M1155$-$79B (see next).

\subsection{SEDs}\label{subsubsec:SED}

In addition to photometry from WISE  \citep{WISE}, archival photometry is available for both 2M1155$-$79B and 2M1155$-$79A from Gaia EDR3 \citep{Gaia2020}, 2MASS \citep{2MASS}, and the Vista Hemisphere Survey \citep{VHS}. The visible and NIR magnitudes were de-reddened for 2M1155$-$79A and 2M1155$-$79B following the same procedure used for the FIRE spectra (see Sec~\ref{sec:obs}). The resulting spectral energy distributions (SEDs) generated from these archival data are presented in Fig.~\ref{fig:SED}. Throughout the optical and mid-IR SED, there is a factor of $\sim$100 luminosity difference between A and B. This is discrepant with the near-IR flux ratio seen in the FIRE data ($\sim$10). Although the flux ratio measured in the FIRE spectra may be unreliable, due to possible chromatic effects from slit placement as well as possible saturation in the J-band portions of the spectra of 2M1155$-$79A and the (A0) flux calibration standard, the large discrepancy between the 2MASS flux ratios and the FIRE spectral ratio is indicative of possible variable obscuration of 2M1155$-$79B (see below). 

The W1-W2 color of 2M1155$-$79A ($0.23 \pm 0.03$) is consistent with that expected from photospheric emission for a star in its (mid-M) spectral type range \citep[i.e., for M5 and M6, W1-W2  = 0.21 and 0.27, respectively;][]{Pecaut2013}. Additionally,  the W4 (22 $\mu$m) flux arises primarily from 2M1155$-$79B (Sec~\ref{subsubsec:WISE}). Therefore, we find no evidence for a warm circumstellar dust component associated with 2M1155$-$79A.

\section{Discussion}\label{sec:discussion}

\subsection{The Two Components of 2M1155-79}\label{subsec:compA}

As noted in Sec~\ref{subsec:spec}, the near-IR SED revealed by the FIRE spectra of 2M1155$-$79B is strikingly similar to --- albeit significantly fainter than --- that of 2M1155$-$79A (Fig.~\ref{fig:FIREspec}). In addition to the large flux ratio, a glaring distinction between the two spectra is the presence of the He {\sc i} 1.083 $\mu$m emission line in 2M1155$-$79B. The WISE analysis reveals that the IR excess originally associated with 2M1155$-$79A, instead arises from 2M1155$-$79B (see Sec.~\ref{subsubsec:WISE}). This, combined with the He {\sc i} emission, constitute strong evidence for the presence of a disk surrounding, and accreting onto, 2M1155$-$79B. Edge-on disks appear unusually faint in optical and NIR bands compared to the expected luminosities of a given spectral type, while still presenting an IR excess at longer wavelengths. This is because the stellar photosphere is occulted, such that photospheric radiation that is scattered off of the disk surface dominates the optical and NIR SED (accounting for the diminished flux in that regime), while thermal IR emission from the disk dust still emerges \citep[e.g.,][]{DAlessio2006,Furlan2011}. The very low apparent luminosity and red Gaia/2MASS colors of 2M1155$-$79B \citep{DickVand2020} hence could be due to obscuration of the stellar photosphere by this orbiting, dusty accretion disk. 

\begin{figure}
    \centering
    \includegraphics[width=0.95\linewidth]{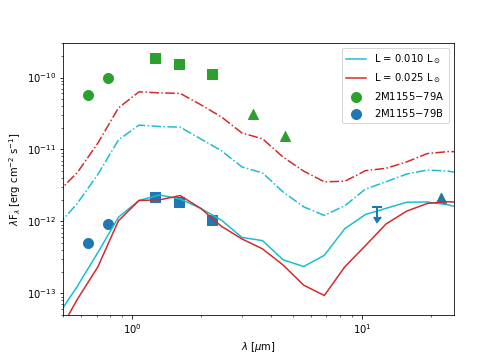}
    \caption{The SED of 2M1155$-$79B (blue points) with two MCFOST models overplotted; the SED of 2M1155$-$79A (green points) are also plotted as a reference. tCircles represent fluxes from Gaia EDR3 (G and G$_{RP}$), squares represent fluxes from 2MASS ($J$, $H$, and $K$), and triangles represent fluxes from WISE (W1, W2, W3, and W4). The $J$ and $K$ fluxes for 2M1155$-$79B are from the Vista Survey. The WISE fluxes are partitioned according to which object clearly dominates in each band; the downward arrow represents the upper limit on the W3 flux for 2M1155$-$79B, since it is not clear how much this source contributes to the emission in this WISE band  (see Sec~\ref{subsubsec:WISE}). The cyan line is the highly-inclined model corresponding to a stellar luminosity of 0.010 L$_\odot$ and the red line is the highly-inclined model corresponding to 0.025 L$_\odot$. The dash-dot lines correspond to the respective pole-on models for 0.010 and 0.025 L$_\odot$. }
    \label{fig:SED}
\end{figure}

As an initial investigation of this hypothesis, we used the radiative transfer modeling code MCFOST \citep{Pinte2006,Pinte2009} to generate two models for the combined star and disk system of 2M1155$-$79B that can reproduce its double-peaked optical through mid-IR SED, as illustrated in Fig.~\ref{fig:SED}. These models represent two extremes of stellar luminosity, 0.010 L$_\odot$ and 0.025 L$_\odot$, with the stellar effective temperature fixed at 3000 K using the \citet{Baraffe2015} stellar spectra models, corresponding to the expected range of stellar properties for $\sim$5 Myr-old M5/M6 pre-MS stars. For both scenarios, we utilized a tapered-edge disk model composed of astronomical silicates --- separated into small (0.01--7 $\mu$m) and large (7--3000 $\mu$m) grains --- with a flaring exponent of 1.07 and a disk scale height of 7 AU at a reference radius of 100 AU \citep{Pinte2009}.
We fixed the large grain dust disk mass at 2.0$\times10^{-5}$ $M_\odot$ and the small grain dust disk mass at 1.5$\times10^{-6}$ $M_\odot$. The small dust grain component remained the same between both models with an inner radius of 0.015 AU and outer radius of 60 AU. To match the observed SED, we find that the 0.010 L$_\odot$ model requires a disk inclination of 75$^\circ$ and an inner radius of 0.2 AU for the large dust grains, while the 0.025 L$_\odot$ model requires an inclination of 81$^\circ$ and an inner radius of 0.45 AU for the large dust grains. We also display the corresponding models for an inclination of 0$^\circ$ (i.e., pole-on) in Fig.~\ref{fig:SED}; these models support the conclusion that obscuration by the disk is responsible for the observed weak stellar photospheric emission signature in the 2M1155$-$79B SED. These initial MCFOST modeling results (Fig.~\ref{fig:SED}) thus lend strong support to the scenario wherein the observed SED of 2M1155$-$79B results from a highly inclined circumstellar disk. 

In contrast, given the WISE data (and lack of He {\sc i} emission from 2M1155$-$79A), we can now conclude that there is no evidence for a circumstellar disk around 2M1155$-$79A. The NIR spectra (Fig.~\ref{fig:FIREspec}) indicate that both 2M1155$-$79A and B are mid- to late-M (M5/6) type stars, whereas optical spectroscopy of 2M1155$-$79A previously established the star as spectral type M3 \citep{Kastner2012}.  The discrepancy between these near-IR (M5/6) and optical (M3) spectral classifications would not be without precedent; it has been previously noted that the optical spectral classifications of pre-MS stars can be 3--5 subclasses earlier than their NIR-based classifications \citep[][]{Kastner2015,Pecaut2016}. However, it is possible that the previous determination of an earlier spectral type for 2M1155$-$79A could reflect the relatively limited wavelength range of the spectrum used for classification.

Furthermore, our MCFOST modeling demonstrates that the occultation of 2M1155$-$79B by this disk --- rather than a difference in stellar mass (hence luminosity) --- also most likely accounts for much of the enormous flux difference between 2M1155$-$79B and 2M1155$-$79A throughout the optical and NIR. The strong similarity of the FIRE spectra of 2M1155$-$79A and 2M1155$-$79B, and their factor $\sim$10 difference in flux levels, should translate to very similar 2MASS colors and a systematic difference of $\sim$2.5 mag in the 2MASS photometry. The redder 2MASS $J-H$ color of 2M1155$-$79B, and the overall $\sim$5 mag difference between the two components in the 2MASS data hence strongly suggests that, at the epoch of the 2MASS survey (1998), 2M1155$-$79B was more heavily occulted by its disk than when we observed in 2020. Such variable obscuration is frequently observed in analogous highly inclined star/disk systems (see \S~\ref{subsec:compOthers}). 
On the other hand, the fact that the Gaia EDR3 $G-G_{RP}$ color of 2M1155$-$79B is not as red as in the DR2 photometry ($G-G_{RP}$ $=$ 1.40 and 1.74, respectively) is more likely due due to a spurious measurement of $G_{RP}$ in DR2. Such Gaia color variability would seem to be inconsistent with the very highly inclined (i$\sim$80$^\circ$), optically thick disk invoked in our models; the photospheric component in these models is dominated by scattering, such that the optical/near-IR colors should be saturated and (hence) constant. Additional  photometric monitoring of this system, along the lines of that conducted for TWA 30AB (see next), is clearly warranted.

\subsection{Comparison to Analogous Systems}\label{subsec:compOthers}

Various young, low-mass star/disk systems, some of them also binaries, provide useful points of comparison with the 2M1155$-$79AB system. Two particularly interesting objects for purposes of comparison are two nearby, low-mass pre-MS systems that also display evidence for highly inclined disks: 2MASS J12014343-7835472 (henceforth 2M1201$-$78) and TWA 30B. 2M1201$-$78 is an early M-type star ($\sim$M2.25) that is also a member of the $\sim$5 Myr $\epsilon$CA \citep{Luhman2004,DickVand2021}. As well as hosting a highly inclined disk (i$\sim$84$^\circ$), optical spectroscopy showed some signs of ongoing accretion in the form of emission from lines of He {\sc i}, [S {\sc ii}], and the Ca {\sc ii} triplet, although other signatures were not detected. Modelling by \citet{Fang2013} attribute the lack of some accretion signatures to a sparse inner disk.

A known member of the TW Hya Association ($\sim$8 Myr), TWA 30B is an M-type (M3/M4) pre-MS star that is a wide-separation ($\sim$ 3400 AU) companion to TWA 30A \citep{Looper2010}. Like the 2M1155$-$79AB system, TWA 30B is fainter than its companion TWA 30A in optical and NIR; it has 5 magnitudes fainter than TWA 30A despite having a slighter earlier spectral type. The TWA 30B disk has been detected via its WISE excess emission and ALMA sub-mm continuum emission \citep{Schneider2012,Rodriguez2015}. Optical and NIR spectra of TWA 30B not only show He {\sc i} emission lines, but also multiple forbidden lines (such as [OI], [OII], and [CI]) indicating that the system is actively accreting and likely drives jets \citep{Looper2010}. 

\begin{figure}
    \centering
    \includegraphics[width=0.95\linewidth]{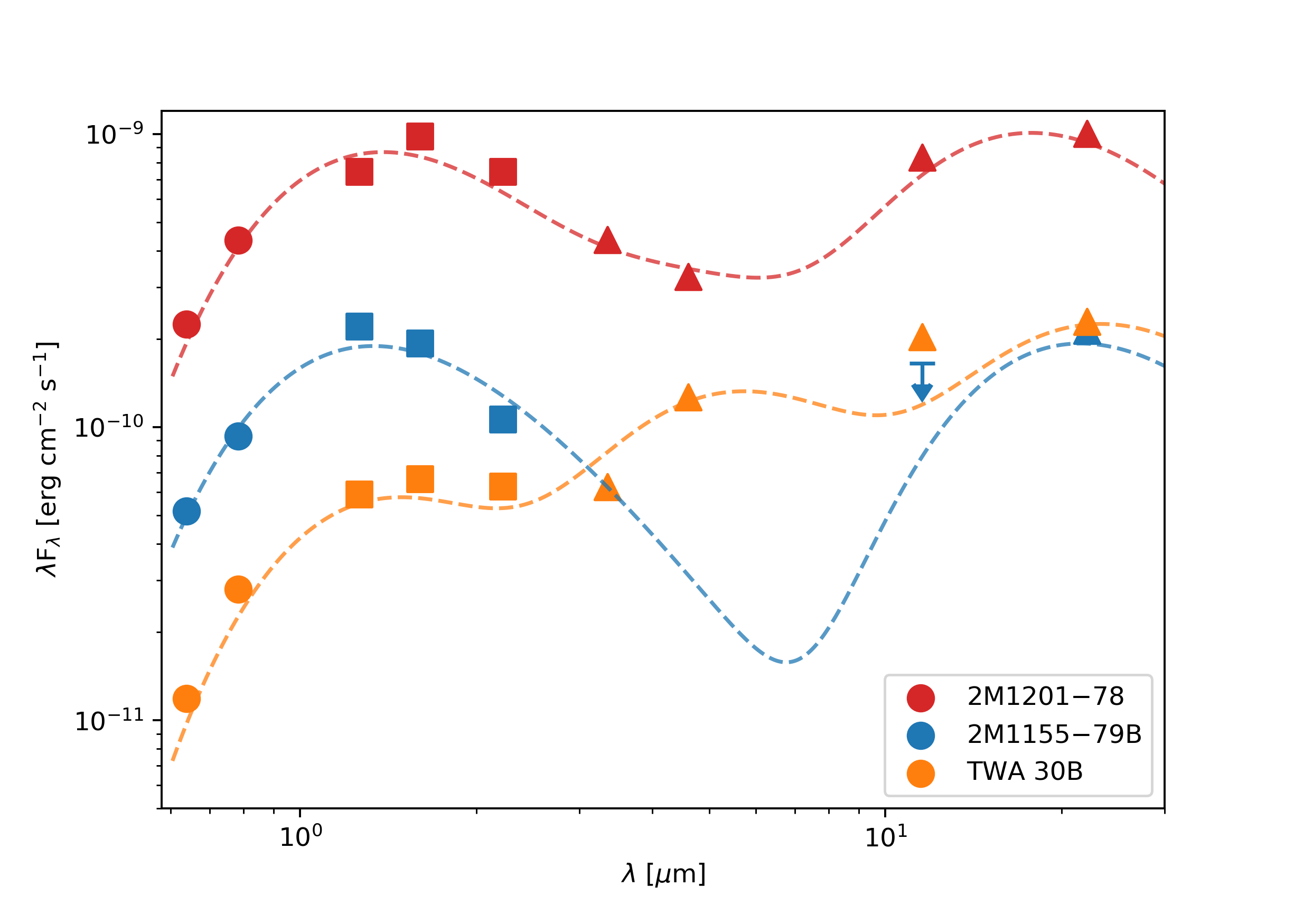}
    \caption{The SEDs of 2M1155$-$79B (in blue), TWA 30B (in orange), and 2M1201$-$78 (in red) displayed on a log-log scale. The data markers are indicative of the same photometric bands as is listed in the caption of Fig~\ref{fig:SED}. The fluxes of all three systems have been normalized to a uniform distance of 10 pc. The dashed lines indicate the combined blackbody model for each system. The stellar photospheres are modeled as simple blackbodies and are not indicative of the actual stellar effective temperatures. The photospheric blackbodies for each system, scaled to the $H$-band flux, are 2750 K for 2M1155$-$79B, 2500 K for TWA 30B, and 2650 K for 2M1201$-$78. The remaining excesses (corresponding to the disks) are modeled using two blackbodies for 2M1201$-$78 ($T = 200, 750$ K) and TWA 30B \citep[$T = 150, 620$ K;][]{Rodriguez2015} and a single blackbody for 2M1155$-$79B ($T = 170$ K).}
    \label{fig:compSEDs}
\end{figure}

In Fig.~\ref{fig:compSEDs}, we compare the SEDs of 2M1155$-$79B, 2M1201$-$78, and TWA 30B, as compiled from data from Gaia EDR3, 2MASS, the Vista Hemisphere Survey, and WISE. The SEDs are overlaid with composite blackbody curves approximating the stars' photospheric and disk emission (dashed lines). Only the photometric data for 2M1155$-$79B have been dereddened; the $\epsilon$CA region toward 2M1201$-$78 appears to show minimal interstellar reddening \citep{Murphy2013}, while the TWA 30B system displays variable reddening due to its dusty circumstellar disk \citep[][]{Looper2010,Principe2016}.

\begin{figure*}
    \centering
    \includegraphics[width=0.95\linewidth]{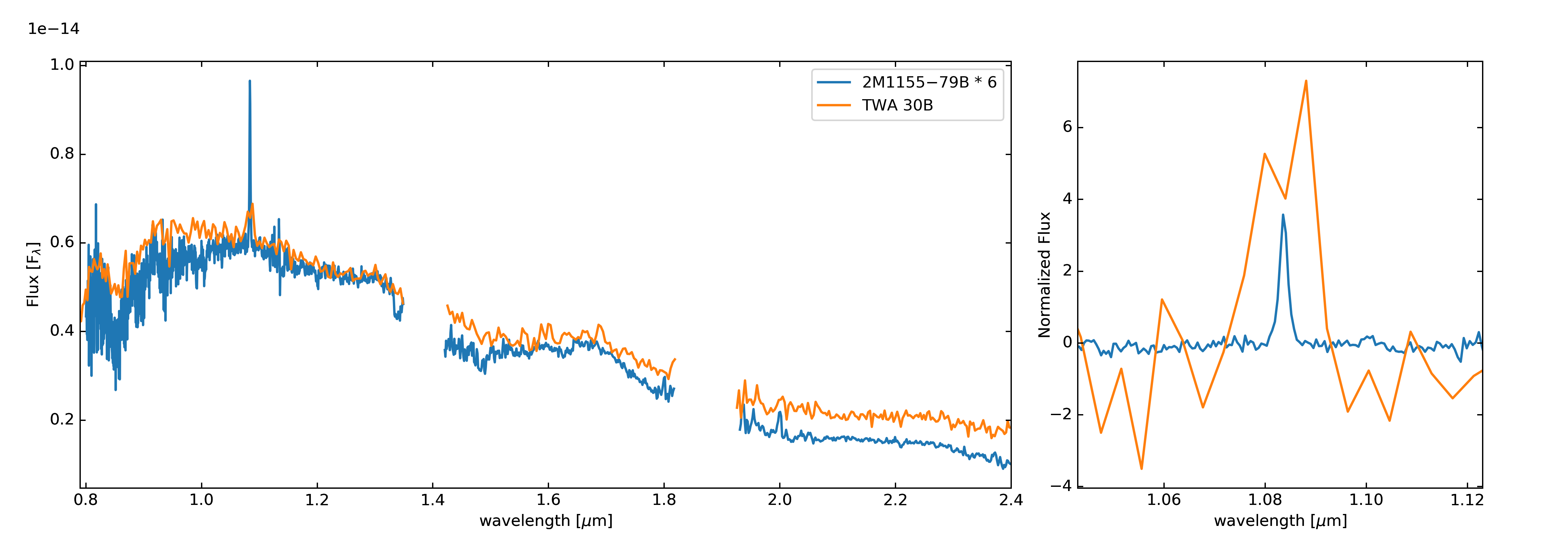}
    \caption{Two IR spectra are displayed in both panels: 2M1155$-$79B (blue, taken with the FIRE instrument) and TWA 30B (orange, taken with the SPeX instrument). In the left panel, the 2M1155$-$79B spectrum has been normalized for visualization with the TWA 30B spectrum by the entire spectrum being multiplied by a constant of 6 and shows the full wavelength coverage. The right panel is centered on the He {\sc i} 1.083 emission line and both fluxes have had the continuum flux subtracted and normalized to 0.}
    \label{fig:compSpecs}
\end{figure*}

While the three objects are similar in nature --- i.e., low-mass stars with large IR excesses --- the comparison of their SEDs indicates there are important differences in either disk structure or disk inclination. TWA30B has a flat SED, indicative of a ``full'' disk with no significant inner cavity, while 2M1201 appears to have a clear break between its photospheric and disk emission, indicative of an inner disk cavity, as hypothesized by \citet{Fang2013}. This is consistent with the accretion signatures from the two objects: 2M1201 has weaker accretion signatures than TWA 30B \citep{Fang2013,Looper2010}. However, small differences in disk inclination could also explain these differences in SEDs. Because of WISE's inability to cleanly resolve 2M1155$-$79B from A, we are unable to establish which of these contrasting SED shapes might best characterize 2M1155$-$79B.

Figure~\ref{fig:compSpecs} compares the NIR spectra of 2M1155$-$79B and TWA 30B \citep[previously classifed as M4;][]{Looper2010,Principe2016}, including the strengths of the He {\sc i} 1.083 $\mu$m emission lines. The TWA 30B IRTF-SpeX spectrum (obtained on 8 June 2011) was among the NIR spectra presented in \cite{Principe2016} least affected by the variable reddening of TWA 30B. The comparison supports our determination of a mid- to late-M spectral type for 2M1155$-$79B (Sec~\ref{subsec:spec}). The 1.083 $\mu$m He {\sc i} line EWs of the two systems are similar: we measure $\sim -$20 \r{A} for TWA 30B, vs.\ $\sim -$12 \r{A} for 2M1155$-$79B (Sec~\ref{subsec:spec}). The line is unresolved in both spectra, so we are unable to compare the linewidths.

As noted in Sec~\ref{subsec:compA}, the apparent optical/near-IR photometric and color variability of 2M1155$-$79B suggests that it experiences variable obscuration by its highly inclined disk; curiously, this behavior more closely resembles that of TWA 30B's wide-separation companion, TWA 30A, than TWA 30B itself \citep{Looper2010,Principe2016}. The disk surrounding TWA 30A is most likely less highly inclined than its companion, as TWA 30A is brighter in the optical/near-IR and, unlike TWA 30B, has been detected in X-rays \citep{Looper2010,Principe2016}.

Another recently discovered low-mass pre-MS star pair in which (like 2M1155$-$79AB) one component harbors a disk and one is diskless is the potential wide binary DENIS-P J1538316-103900 (DENIS1538-1039) and DENIS-P J1538317-103850 (DENIS1538-1038). DENIS1538-1038 is a $\sim$1 Myr-old brown dwarf (spectral type M5.5) with an IR-excess indicative of a disk \citep{N-T2020}. The M3 star DENIS1538-1039  is seen at 10$''$ projected separation from DENIS1538-1038. \citet{N-T2020} conclude that the two stars are potential members of the Upper Scorpius Association, although \citep[as noted by][]{N-T2020}, at an estimated age of $\sim$1--5 Myr, they are much younger than the average age of the association \citep[$\sim$10 Myr,][]{Luhman2020}; they are also far from the main Upper Scorpius Association complex, further casting their membership in doubt. On the basis of Gaia DR2 data, \citet{N-T2020} concluded that the pair represent a chance line of sight alignment of two members of the same NYMG. However, in Gaia EDR3, the two components have the same proper motions and parallaxes within the uncertainties, indicating the two stars are in fact a wide binary analogous to 2M1155$-$79AB, i.e., a wide separation pair of young M-stars near the stellar/sub-stellar boundary, one with a disk and one without. 

Two other systems merit mention here, as potential younger analogs to the 2M1155$-$79AB wide binary. Like 2M1155$-$79B, the companion to the 1--2 Myr-old 2MASS J19005804$-$3645048 was initially thought to be a planet-mass object before follow-up spectroscopy instead suggested that the star was a young late-M dwarf \citep{Christiaens2021}. \citet{Christiaens2021} discuss the implication that the companion is an obscured low-mass star for which only a small fraction of its light emerges due to an edge-on disk, as well as the possibility that the companion is an accreting protoplanet that is being heated by accretion shocks. It is unlikely that 2M1155$-$79B falls into this second category, given the lack of WISE detection of warm dust around 2M1155$-$79A (\S~\ref{subsubsec:WISE}). The second system, HK Tau AB, is a low-mass, wide binary in the Taurus star forming region that, like 2M1155$-$79AB, consists of two stars of similar (in this case, early-M) spectral type, but with B many magnitudes fainter than A \citep{Monin1998}.  NIR imaging of the system revealed that HK Tau B is occulted by an edge-on disk \citep{Staplefeldt1998}. Subsequent scattered light and sub-mm observations demonstrate that both components possess extensive gas and dust disks, but with sharply contrasting inclinations  \citep[i$\sim$43$^\circ$ and i$\sim$85$^\circ$ respectively; ][]{McCabe2011,Jensen2014}.

It is also worth considering whether 2M1155$-$79AB, as well as some of the aforementioned systems, constitute examples of young, hierarchical multiples wherein at least one component harbors a relatively long-lived, dusty disk \citep[][and references therein]{Kastner2018}. There are multiple instances of circumbinary disks in such systems, as well as young hierarchical multiple systems showing the presence of disks around some components and not others; the HD 104237 system within $\epsilon$CA is an example of a particularly complex system with both of these characteristics \citep{Murphy2013,DickVand2021}. Thus far, however, there is no evidence that 2M1155$-$79AB is such a (hierarchical) multiple system \citep[in particular, DR2 photometry are consistent with 2M1155$-$79A being a single star;][]{DickVand2021}. 

\subsection{The Nature of 2M1155-79B}\label{subsec:nature}

While the original evidence in \citet{DickVand2020} pointed towards 2M1155$-$79B being a nascent, 10 M$_{jup}$ planet, the spectral type determined here, M5/6, is inconsistent with this picture. Nonetheless, 2M1155$-$79B is potentially still near the hydrogen burning limit; pre-MS model evolutionary tracks place the stellar versus brown dwarf boundary around a spectral type of M6 at 3-5 Myr \citep{Baraffe2015}. A firm conclusion as to the fate of 2M1155$-$79B --- low-mass star or brown dwarf --- will require optical spectroscopy to confirm the spectral type of this enigmatic system.

To our knowledge, 2M1155$-$79B is the latest M-type pre-MS star in which the He {\sc i} 1.083 $\mu$m emission line has yet been detected. In higher-mass pre-MS (T-Tauri) stars, this line is a sensitive probe of accretion shocks and accretion-driven winds \citep{Kwan2007}, manifested in the form of red- and blueshifted absorption features that are a consequence of the high 1.083 $\mu$m line opacity \citep{Edwards2006,KwanFisher2011,Sousa2021}. For a high-inclination star-disk system with stellar and/or disk winds, the He {\sc i} 1.083 $\mu$m emission has a narrow, blue-shifted absorption feature due to the emission passing through the slow disk wind component \citep{Kwan2007}. Given the resolution of our spectra, we cannot retrieve any information about potential blue-shifted absorption; however, the EW of the emission line from 2M1155$-$79B is similar to that of mid-K classical T Tauri star disks \citep{Edwards2003}, and our initial modeling supports the presence of a nearly edge-on circumstellar disk in the system (Fig~\ref{fig:SED}).  Modeling by \citet{KwanFisher2011} indicates that He {\sc i} emission could be arising from the accretion flow close to a stellar impact shock, resulting in UV photoionization and temperatures of at least 10$^4$K.  However, since 2M1155$-$79B has a much lower mass and therefore should have a lower shock temperature than in the \citet{KwanFisher2011} models, the He {\sc i} 1.083 $\mu$m could be indicative of the presence of a large-scale wind or jet. NIR imaging of this object might reveal a jet origin for the He {\sc i} 1.083 $\mu$m emission, while higher resolution spectroscopy of the He {\sc i} 1.083 $\mu$m emission will allow for analysis of intervening kinematic structures through the study of potential absorption components within the line profile. The large EW of this line in the spectrum of 2M1155$-$79B makes this object a strong candidate for the use of both of these methods.

\section{Conclusions}\label{sec:conclusion}

The object 2M1155$-$79B is a low-mass member of the $\sim$5 Myr moving group, $\epsilon$CA, and a wide separation companion (5.75", 580 AU) to another $\epsilon$CA member, 2M1155$-$79A. The SED of 2M1155$-$79B is $\sim$10--100 times fainter than 2M1155$-$79A throughout the optical and NIR (Figs.~\ref{fig:FIREspec}, \ref{fig:SED}). The extreme faintness and redness of 2M1155$-$79B in Gaia DR2 and archival 2MASS photometry presented the scenario that 2M1155$-$79B was a planet-mass object (M$\sim$10 M$_{jup}$). However, we have obtained near-infrared spectroscopy with Magellan/FIRE demonstrating that the object is best matched by a young star standard spectra in the range M5--M6 (Fig.~\ref{fig:FIREspec}). The fact that 2M1155$-$79B is much fainter in the optical and NIR than expected of a star of this spectral type is hence most likely due to partial occultation of the stellar photosphere by a highly inclined circmstellar disk. Furthermore, a strong He {\sc i} 1.083 $\mu$m emission line is observed in the FIRE spectra of 2M1155$-$79B, indicative of ongoing accretion and accretion-driven winds and/or jets (see Sec~\ref{subsec:nature}). Nevertheless, the FIRE spectra of 2M1155$-$79B and 2M1155$-$79A are very similar in shape, revealing that the two stars may be near-twins, both with spectral types in the range M5--M6. The spectral type we infer for 2M1155$-$79A from its FIRE spectrum is somewhat later than that previously determined from its optical spectrum. Follow-up optical spectroscopy of 2M1155$-$79B is needed to clarify the photospheric properties of 2M1155$-$79B and verify that the two components of the 2M1155$-$79AB binary are indeed near-twins.
Analysis of the WISE photometry for the 2M1155$-$79AB system (Sec~\ref{subsubsec:WISE}, Fig.~\ref{fig:WISEcent}) demonstrates that the infrared excess originally associated with 2M1155$-$79A instead originates from 2M1155$-$79B, supporting the  interpretation that the secondary is orbited and occulted by a dusty disk. Modeling of the SED of the 2M1155$-$79B star-disk system (Fig.~\ref{fig:SED}) show that a highly inclined disk (i$\sim$75-81$^\circ$) can account for the flux difference between primary  and secondary.  

As discussed in Sec~\ref{subsec:compOthers}, these results place 2M1155$-$79B among a small subset of pre-MS, low-mass stars that are both highly obscured by edge-on disks and show signs of active accretion. Like these other, analogous systems, the 2M1155$-$79B system is a particularly promising subject for studies of star and planet formation around low-mass stars.

\acknowledgements

We thank the anonymous referee for helpful comments and suggestions. This research is supported by NASA Exoplanets Program grant 80NSSC19K0292 to Rochester Institute of Technology. This work has made use of data from the European Space Agency (ESA) mission {\it Gaia} (\url{https://www.cosmos.esa.int/gaia}), processed by the {\it Gaia} Data Processing and Analysis Consortium (DPAC, \url{https://www.cosmos.esa.int/web/gaia/dpac/consortium}). Funding for the DPAC has been provided by national institutions, in particular the institutions participating in the {\it Gaia} Multilateral Agreement. This publication makes use of data products from the Two Micron All Sky Survey, which is a joint project of the University of Massachusetts and the Infrared Processing and Analysis Center/California Institute of Technology, funded by the National Aeronautics and Space Administration and the National Science Foundation. This publication makes use of data products from the Wide-field Infrared Survey Explorer, which is a joint project of the University of California, Los Angeles, and the Jet Propulsion Laboratory/California Institute of Technology, funded by the National Aeronautics and Space Administration.

\bibliography{main}{}
\bibliographystyle{aasjournal}

\end{document}